\documentclass[aps,prl,reprint,superscriptaddress]{revtex4-2}
\usepackage[breaklinks=true,colorlinks,citecolor=blue,linkcolor=blue,urlcolor=blue]{hyperref}

\usepackage{amssymb,amsmath,amsfonts}
\usepackage{subfigure, latexsym,multirow}
\usepackage[notcite,color,final]{showkeys}
\usepackage{epsfig,graphicx,subfigure,dcolumn,bm}
\usepackage[usenames ,dvipsnames]{xcolor}
\usepackage{slashed}
\usepackage{ulem}
\usepackage{amsfonts}

\usepackage{array}  

\usepackage{ulem}
\usepackage{booktabs}

\newcommand{\addtext}[1]{\textcolor{black}{#1}}
\begin{document}

\title{\addtext{ Hidden Quantum-Geometric Connection between Injection and Shift Currents}\textsf{}}

\author{Mohammad Yahyavi}
\affiliation {Division of Physics and Applied Physics, School of Physical and Mathematical Sciences, Nanyang Technological University, 21 Nanyang Link 637371, Singapore}
 
\author{Tay-Rong Chang}
\affiliation {Department of Physics, National Cheng Kung University, Tainan, Taiwan}
\affiliation {Center for Quantum Frontiers of Research and Technology (QFort), Tainan, Taiwan}
\affiliation {Physics Division, National Center for Theoretical Sciences, Taipei 10617, Taiwan}

\author{Md Shafayat Hossain}
\affiliation{Department of Materials Science and Engineering, University of California, Los Angeles, California 90095, USA}
\affiliation{California NanoSystems Institute, University of California, Los Angeles, California 90095, USA}
\affiliation{Center for Quantum Science and Engineering, University of California, Los Angeles, California, USA}

\author{Arun Bansil}
\affiliation{Department of Physics, Northeastern University, Boston, MA 02115, USA}
\affiliation{Quantum Materials and Sensing Institute, Northeastern University, Burlington, MA 01803, USA}

\author{Naoto Nagaosa}
\affiliation {RIKEN Center for Emergent Matter Science (CEMS), Wako, Saitama 351-0198, Japan}
\affiliation {Fundamental Quantum Science Program (FQSP), TRIP Headquarters, RIKEN, Wako 351-0198, Japan}

\author{Guoqing Chang}
\email{guoqing.chang@ntu.edu.sg}
\affiliation {Division of Physics and Applied Physics, School of Physical and Mathematical Sciences, Nanyang Technological University, 21 Nanyang Link 637371, Singapore}


 \begin{abstract}
	{ \addtext{Injection current (IC) and shift current (SC) are generally regarded as distinct nonlinear optical responses with different microscopic origins. Here, within the clean limit, based on an independent-particle interband description, we uncover a general hidden connection between the IC and SC responses. The generalized-derivative sum rule is used to show that the SC response separates into an injection-determined contribution associated with velocity asymmetry and a genuine shift-current (GSC) contribution associated with real-space wave-packet displacement. In systems with approximately linear electronic dispersion near the Fermi level, the corresponding tensor combinations of SC and IC can become effectively equivalent at low photon energies, governed by the same interband quantum-geometric dipole. Crystal symmetry can also eliminate the injection-determined contribution, enabling direct access to the GSC component. Extending our approach to second-harmonic generation reveals analogous response relations, suggesting a route toward a broader network of quantitative connections among the various nonlinear optical responses.}}
	
\end{abstract}


\maketitle
Uncovering hidden connections among seemingly distinct physical effects has often led to fundamental advances in physics. Such connections can emerge as effective equivalences, whereby different phenomena become indistinguishable within an appropriate theoretical framework \cite{Gibbs1902,Einstein1905,His1,His2}. In optics, geometrical optics arises as an effective description of wave optics in the short-wavelength limit. Beyond the linear regime, nonlinear optical responses provide a powerful framework for light--matter interactions and have become important probes of symmetry, topology, and quantum geometry in quantum materials~\cite{Nonlinear_Book1,Sipe2000,Thz2,NH-03,Nonlinear__perspective,Nagaosa17,Nonlinear_Book_2,Thz_3, NH-01,NH-02,NH-08,Ma2021Topology,Bao2022,NH-09,BCD,SHG_P1,Morimoto16,NH-04,chan2017photocurrents,NH-05,NH-06,de2017quantized1,chang2017unconventional1,NH-07,Chang2018,Flicker2018,SC2,SHG_P2,Juan_2020,SHG_P3,Nagaosa2020Low,Ahn2021,Thz_2,QMD,NH-10,NH-11,NH-12,PRL_CPGE,Fregoso2017}. Yet they are commonly classified as distinct phenomena, associated with different microscopic mechanisms and polarization selection rules.

The injection current (IC) and shift current (SC), two prototypical second-order photocurrents, are widely regarded as fundamentally different responses~\cite{Sipe2000,Morimoto16,Nagaosa17,SC2,Fregoso2017,chan2017photocurrents,Nagaosa2020Low,Ahn2021,NH-08,NH-09}. The IC originates from asymmetries in carrier velocities in momentum space, whereas the SC arises from real-space displacements of electronic wave packets during photoexcitation. Their distinction is further reinforced by polarization selection rules: in time-reversal ($\mathcal{T}$)-symmetric systems, the IC is induced by circularly polarized light, whereas the SC is driven by linearly polarized light; in magnetic systems preserving $\mathcal{PT}$ symmetry, this correspondence is reversed. Whether such distinctions reflect genuinely independent physical processes or conceal a deeper \addtext{connection} remains unclear.

\addtext{Here, based on the generalized-derivative sum rule~\cite{Sipe2000,Nagaosa2020Low,Ahn2021}, we reveal a hidden connection between IC and SC. We decompose SC into an IC-determined contribution associated with velocity asymmetry and a genuine shift-current (GSC) contribution associated with real-space wave-packet displacement. In Dirac/Weyl semimetals, as well as in strained twisted bilayer graphene (TBG), IC and SC can become effectively equivalent, with corresponding tensor combinations governed by the same underlying quantum-geometric dipole. Crystal symmetry can also force the IC-determined contribution to vanish, allowing the measured SC to directly probe the GSC contribution. Our approach can be further extended to second-harmonic generation (SHG), providing a systematic route to deriving quantitative relations between the various nonlinear optical responses.}

The second-order photocurrents induced by light of frequency $\omega$ can be expressed as, $J^a = \sigma_{\text{L}}^{abc}(\omega)\,
\text{Re}\left[E^b(\omega)\, E^c(-\omega)\right] +   \sigma_{\text{C}}^{abc}(\omega)\,\text{Im}\left[E^b(\omega)  E^c(-\omega)\right]$, with $\sigma^{abc}_{\text{L}}(\omega)$ [$\sigma^{abc}_{\text{C}}(\omega)$] denoting the second-order dc conductivity tensor associated with linearly polarized (circularly polarized) light. The first tensor index $a$ specifies the direction of the induced current, while the remaining indices $b$ and $c$ correspond to the polarization directions of the electric field. \addtext{We consider resonant interband photocurrents within the independent-particle description in the clean limit, where the optical broadening is small compared to the photon energy and the band gap. In this limit, the dc response consists of two contributions, the SC and the IC~\cite{Sipe2000,Nagaosa2020Low}.}

For IC, the explicit forms of the conductivity tensors for linearly   ($\sigma_{\text{LIC}}$) and circularly-polarized  ($\sigma_{\text{CIC}}$) light    \cite{Sipe2000,Ahn2021}  (Supplemental Material (SM) Sec. I) are:
\begin{equation}
	\begin{split}
	\sigma_{\text{LIC}}^{abc}(\omega)
&= \tau\frac{\pi e^3}{\hbar^2} 
\int_{\mathbf{k}} \sum_{n,m}
f_{nm}\,\Delta_{mn}^a \left\{r_{nm}^b, r_{mn}^c\right\}
\delta(\omega_{mn} - \omega),   \\
\sigma_{\text{CIC}}^{abc}(\omega)
&= -\tau\frac{i\pi e^3}{\hbar^2}
\int_{\mathbf{k}} \sum_{n,m}
f_{nm}\,\Delta_{mn}^a \left[ r_{nm}^b, r_{mn}^c \right]
\delta(\omega_{mn} - \omega),  
\label{eq:dc}
	\end{split}
\end{equation}
where $\int_{\mathbf{k}} = \int d^d k/(2\pi)^d$,  $\hbar \omega_{nm}=\hbar \omega_{n}-\hbar \omega_{m}$ is the energy difference between bands $n$ and $m$ and $f_{nm}=f_{n}-f_{m}$ with $f_{n}$ denoting the Fermi-Dirac distribution of the band $n$. $\Delta_{mn}^c=\nu_{mm}^c-\nu_{nn}^c$ with $\nu_{nm}^c= \langle  u^n_{\mathbf{k}}|(\partial_{k_c} H)/\hbar| u^m_{\mathbf{k}}\rangle$ refer to intra-band (inter-band) velocity when $m=n$ ($m\neq n$). $r^{b}_{nm} = i\langle u^n_{\mathbf{k}} |\partial_{k_b}| u^m_{\mathbf{k}} \rangle$ is the interband Berry connection, which is in general complex. Its symmetric part $\left\{r_{nm}^b, r_{mn}^c\right\}$ is the quantum metric, which is real, while its antisymmetric part corresponds to the Berry curvature  $\left[ r_{nm}^b, r_{mn}^c \right]$ and it is purely imaginary.  \addtext{ Here, $\tau$ is a commonly invoked phenomenological current-relaxation time used to express the steady-state injection conductivity. In the absence of relaxation, the injected current grows linearly in time within the perturbation theory, with an injection-rate coefficient given by $\sigma_{\mathrm{IC}}/\tau$.}

Conductivity tensors  of SC induced by linearly polarized light ($\sigma_{\text{LSC}}$) and circularly-polarized light ($\sigma_{\text{CSC}}$)  are (SM Sec.  I) \cite{Sipe2000,Ahn2021}:
\begin{equation}
	\begin{split}
		\sigma_{\text{LSC}}^{abc}(\omega)
		&= -\frac{i\pi e^3}{2\hbar^2}
		\int_{\mathbf{k}} \sum_{n,m}
		f_{nm}\,   \mathcal{I}_{mn}^{abc}\;
		D_{+}(\omega_{mn},\omega), \\
		\sigma_{\text{CSC}}^{abc}(\omega)
		&= \frac{\pi e^3}{2\hbar^2}
		\int_{\mathbf{k}} \sum_{n,m}
		f_{nm}\,  \mathcal{R}_{mn}^{abc}\;
		D_{-}(\omega_{mn},\omega).
		\label{eq:dc}
	\end{split}
\end{equation}
where $\mathcal{I}_{mn}^{abc}=r_{mn}^b r_{nm;a}^c+ r_{mn}^c r_{nm;a}^b$, $\mathcal{R}_{mn}^{abc}=r_{mn}^b r_{nm;a}^c- r_{mn}^c r_{nm;a}^b$, and $D_{\pm}(\omega_{mn},\omega)=\delta(\omega_{mn}-\omega)\pm\delta(\omega_{mn}+\omega)$. The covariant derivative $r_{nm;a}^c = C_{nm}^{c,a} \, r_{nm}^c$ contains the ``shift vector'' $C_{nm}^{c,a} = \partial_a \log r_{nm}^c - i (A_n^a - A_m^a)$, which characterizes the real-space displacement during interband optical transitions, with $A_{n}^a      =i\langle  u^n_{\mathbf{k}}|\partial_{k_a}| u^n_{\mathbf{k}}\rangle$ denoting the intraband Berry connection. 

 \addtext{Applying the generalized-derivative sum rule, we resolve $\mathcal{I}_{mn}^{abc}$ ($\mathcal{R}_{mn}^{abc}$)} into three contributions (SM Sec.  I.A): (i) terms arising from the first-order derivative of the Hamiltonian and wavefunctions $\mathcal{I}_{mn}^{^1 abc}$ ($\mathcal{R}_{mn}^{^1 abc}$); (ii) terms involving the second-order derivative of the Hamiltonian $\mathcal{I}_{mn}^{^2 abc}$ ($\mathcal{R}_{mn}^{^2 abc}$); and (iii) terms associated with interband virtual transitions $\mathcal{I}_{mn}^{^3 abc}$ ($\mathcal{R}_{mn}^{^3 abc}$). 

The first contribution $\mathcal{I}_{mn}^{^1abc}$ to the linearly polarized-light induced shift current (LSC)  \cite{Sipe2000} is:
\begin{equation}
	\begin{split}
		\mathcal{I}_{mn}^{^1 a b c}
		=
		\frac{\Big(
			\Delta_{mn}^{c}\, r_{mn}^{b} r_{nm}^{a}+
			\Delta_{mn}^{b}\, r_{mn}^{c} r_{nm}^{a}
			+
			\Delta_{mn}^{a}\, \{ r_{mn}^{b}, r_{nm}^{c} \}
			\Big)}{\omega_{nm}}
		,
	\end{split}
\end{equation}
Splitting $r_{mn}^{b} r_{nm}^{a}$ into $\left(\{ r_{mn}^b, r_{nm}^a \}+[ r_{mn}^b, r_{nm}^a ]\right)/2$, we can rewrite $\mathcal{I}_{mn}^{^1abc}$ as:
\begin{equation}
	\begin{split}
		\mathcal{I}_{mn}^{^1 a b c}
		=&
		\frac{1}{2\omega_{nm}}
		\Big(
		\Delta_{mn}^c \{ r_{mn}^b, r_{nm}^a \}
		+
		\Delta_{mn}^b \{ r_{mn}^c, r_{nm}^a \}\\
		&+
		2\Delta_{mn}^{a}\, \{ r_{mn}^{b}, r_{nm}^{c} \}
		\\&+\Delta_{mn}^c [ r_{mn}^b, r_{nm}^a ]
		+
		\Delta_{mn}^b [ r_{mn}^c, r_{nm}^a ]	\Big).
	\end{split}
\end{equation}
where the first three symmetrized products give the real part, while the last two antisymmetrized products give the imaginary part.

\begin{figure} 
	\includegraphics[width=\linewidth]{main/Fig1.pdf}
	\caption{\addtext{{\bf Emergent quantum-geometric equivalence between IC and SC in Dirac/Weyl semimetals.}
		(a) LSC  of a linear Weyl node model and
				(b) LSC of TaAs from first-principles band structure.
		Red dashed lines correspond to the analytical results obtained from our formula [Eq.~(\ref{eq:11})], while the solid blue lines  show the numerical evaluation of the linear shift conductivity from Eq.~(\ref{eq:dc}).
		(c) Schematic of the combined polarization measurement, moiré
		Brillouin zones, and band-energy dispersion at the $K^{+}$ valley of strained TBG for a twist angle of $\theta = 10^\circ$. Green and red arrows mark the incident photon energy $\hbar\omega_{i}$ and the virtual-transition energy $\hbar\omega_{V}$ to the remote bands, respectively.
				(d) Top: linear shift conductivity $\sigma_{\mathrm{LSC}}^{yxx}$ in strained TBG from the injection-current-based expression (red dashed line) and the full numerical calculation (solid blue line). Bottom: relative deviation $\eta_{v}$ due to virtual transitions as a function of $\hbar\omega_{i}/\hbar\omega_{V}$; it remains below 3\% over the plotted range. 
		(e)  Brillouin zone, crystal structure, and band structure of MnGeO$_3$ along high-symmetry directions. 
		(f) CSC $\sigma_{\mathrm{CSC}}^{xzx}$ in MnGeO$_3$.
		The red dashed line gives the prediction based on the LIC,
		$\sigma_{\mathrm{CSC}}^{xzx} \propto (\sigma_{\mathrm{LIC}}^{xzx}-\sigma_{\mathrm{LIC}}^{zxx})/(2\tau\omega)$,
		while the solid blue line shows the full numerical result. (g) Left: $\sigma_{\mathrm{CIC}}^{zxy}$ probes $D^{z;xy}$, where the Berry-curvature-dipole direction is aligned with the direction of the current.  Right:   $\sigma_{\mathrm{LSC}}^{zxy}$ detects $D^{x;yz}$ and $D^{y;xz}$, where the Berry-curvature-dipole direction is aligned with the polarization direction of the light that drives the SC.}}
	\label{Fig.1}
\end{figure} 

{{Although $\mathcal{I}_{mn}^{^1 a b c}$ is generally complex, its contribution
		to the LSC enters exclusively through the symmetric
		frequency combination
		$D_{+}(\omega_{mn},\omega)
		=\delta(\omega_{mn}-\omega)+\delta(\omega_{nm}-\omega)$.
		Under the exchange of band indices $m\leftrightarrow n$,
		one has $\mathcal{I}_{mn}^{^1 a b c}\rightarrow
		\mathcal{I}_{nm}^{^1 a b c}=(\mathcal{I}_{mn}^{^1 a b c})^{*}$,
		$f_{nm}\rightarrow -f_{nm}$, and
		$\omega_{mn}\rightarrow -\omega_{mn}$, while
		$D_{+}(\omega_{mn},\omega)$ remains invariant.
		As a result, the real (symmetrized) part of
		$\mathcal{I}_{mn}^{^1 a b c}$ is even under band exchange and therefore
		cancels identically upon summation over $m,n$.
		In contrast, the imaginary (antisymmetrized) part is odd under
		$m\leftrightarrow n$ and survives.
		Consequently, the product
		$\mathcal{I}_{mn}^{^1 a b c}D_{+}(\omega_{mn},\omega)$ is purely imaginary,
		ensuring that the linear shift-current conductivity
		$\sigma_{\mathrm{LSC}}^{abc}$ is strictly real, and one finds (see SM Sec. II)}}
\begin{equation}
	\begin{split}
		\mathcal{I}_{mn}^{^1 ab c}D_{+}(\omega_{mn},\omega)=&\frac{1}{\omega_{nm}} \,\left(\Delta_{mn}^c\left[r_{mn}^b\,,r_{nm}^a\right] \right.\\& \left.+\, \Delta_{mn}^b\left[r_{mn}^c\,,r_{nm}^a\right]\,\right) \delta(\omega_{mn}-\omega).
	\end{split}
\end{equation}
This result implies that, for LSC, the contribution arising from the first-order derivative of the Hamiltonian can be expressed as a linear combination of the conductivity tensor of the circularly-polarized-light induced injection currents (CIC): $\sigma_{\text{LSC-I}}^{abc}= \frac{1}{2\tau\omega} \left(\sigma^{cba}_{\text{CIC}}+ \sigma^{b ca}_{\text{CIC}} \right)\label{EQ_5}$.
And thus the full conductivity tensor of LSC can be written as:
\begin{equation}
	\begin{split}	 
		\sigma_{\text{LSC}}^{abc}=& \frac{1}{2\tau\omega} \left(\sigma^{cba}_{\text{CIC}}+ \sigma^{b ca}_{\text{CIC}} \right)+\sigma_{\text{LSC-}\mathcal{W}}^{abc} + \sigma_{\text{LSC-}\mathcal{V}}^{abc} ,  \\
	\end{split}	\label{eq:SC-IC_dec}
\end{equation}
where $\sigma_{\text{LSC-}\mathcal{W}}^{abc} =-\frac{i\pi e^3}{2\hbar^2}  \int_{\textbf{k}}  \sum_{n,m}  f_{nm}\;   \mathcal{I}_{mn}^{^2 ab c} \;  D_+(\omega_{mn},\omega)$, and $\sigma_{\text{LSC-}\mathcal{V}}^{abc} =-\frac{i\pi e^3}{2\hbar^2}  \int_{\textbf{k}}  \sum_{n,m}  f_{nm}\;   \mathcal{I}_{mn}^{^3 ab c} \;  D_+(\omega_{mn},\omega)$.

	 \addtext{ The same decomposition links circularly polarized shift current (CSC) to linearly polarized injection current (LIC)
(SM Sec. III):}
	 
	\begin{equation}
		\sigma_{\text{CSC}}^{abc} =
		\frac{-1}{2\tau\omega}
		\left(\sigma^{cba}_{\text{LIC}} - \sigma^{bca}_{\text{LIC}}\right)
		+ \sigma_{\text{CSC-}\mathcal{W}}^{abc}
		+ \sigma_{\text{CSC-}\mathcal{V}}^{abc}.
		\label{eq:CSC-LIC2}   
	\end{equation}
	where $\sigma_{\text{CSC-}\mathcal{W}}^{abc} =\frac{\pi e^3}{2\hbar^2}  \int_{\textbf{k}}  \sum_{n,m}  f_{nm}\;   \mathcal{R}_{mn}^{^2 ab c} \;  D_-(\omega_{mn},\omega)$, and $\sigma_{\text{CSC-}\mathcal{V}}^{abc} =\frac{\pi e^3}{2\hbar^2}  \int_{\textbf{k}}  \sum_{n,m}  f_{nm}\;   \mathcal{R}_{mn}^{^3 ab c} \;  D_-(\omega_{mn},\omega)$.

	\addtext{Equations (6) and (7) are exact within the clean limit interband framework and hold for arbitrary band structures; no linearization or low-frequency assumption has been made.} These results demonstrate that IC and SC, despite originating from apparently distinct physical mechanisms, are linked by a well-defined functional relation. \addtext{We next discuss the physical interpretation and possible experimentally testable consequences based on the above connections.}

%
	\addtext{\bf{Emergent quantum-geometric equivalence between IC and SC.}} Note that $\sigma_{\text{SC-}\mathcal{W}}^{abc}$ originates from the second-order derivative of the Hamiltonian. Therefore, if the band dispersion is approximately linear,   $\sigma_{\text{SC-}\mathcal{W}}^{abc}$ would be negligible (SM Sec. IV.A). Suppressing $\sigma_{\text{SC-}\mathcal{V}}^{abc}$, on the other hand, requires the frequency of virtual interband transitions associated with high-energy bands ($\omega_{V}$) to be much larger than the incident photon frequency ($\omega_{i}$). When both conditions are satisfied, within a linearized band approximation in the low-frequency limit ($\omega_{i} \ll \omega_{V}$), the right-hand side of Eqs. (\ref{eq:SC-IC_dec}) and  (\ref{eq:CSC-LIC2})    reduces to just the IC contribution: 
	\begin{equation}
		\begin{split}
			\sigma_{\text{LSC}}^{abc} \approx& \frac{1}{2\tau\omega} \left(\sigma^{cba}_{\text{CIC}}+ \sigma^{b ca}_{\text{CIC}} \right), \,\;\;\; \sigma_{\text{CSC}}^{abc} \approx \frac{-1}{2\tau\omega}\left(\sigma^{cba}_{\text{LIC}} - \sigma^{bca}_{\text{LIC}}\right).\\
		\end{split}\label{eq:11}
	\end{equation}
Natural platforms for testing our theoretical framework are the massless Dirac and Weyl semimetals. In this connection, we first consider a two-band model of a Weyl node, in which virtual transitions to remote bands
are absent. Moreover, when only the velocity and tilt terms are retained, the second-order derivative of the Hamiltonian vanishes. Taking $\sigma_{\text{LSC}}^{zxx}$ as a representative quantity, our numerical results show that it equals $\frac{1}{\tau\omega}\sigma^{xxz}_{\text{CIC}}$ [Fig. \ref{Fig.1}(a)].

\addtext{To quantify the correction from second derivatives of the Hamiltonian, we include a finite warping term (SM Sec.~IV.B) and
define $\eta\equiv\left|\sigma_{\text{LSC}}^{abc}-\frac{1}{2\tau\omega}\left(\sigma^{cba}_{\text{CIC}}+\sigma^{bca}_{\text{CIC}}\right)\right|/\left|\sigma_{\text{LSC}}^{abc}\right|$, for a nonzero denominator. Since this model has only two bands, $\eta$ measures the second-derivative correction alone. Although $\eta$ increases with the strength of this term, the second-derivative correction remains below 10$\%$ for the TaAs family [Fig.~\ref{Fig.1}(b) and SM Fig. S2]. }

\addtext{To quantify the deviation caused by virtual transitions, we use a model of strained TBG (SM Sec.  IV.C and  Sec.  IV.D).} We numerically calculated the IC and SC in strained TBG at a twist angle of $\theta = 10^\circ$, where the linear dispersion near the $K$ points remains well preserved while the inversion symmetry ($\mathcal{P}$) is broken [Fig.~\ref{Fig.1}(c)]. \addtext{For photon frequencies well below those of virtual transitions to the remote bands ($\omega_{i} \ll \omega_{V}$), the deviation $\eta$ resulting from the virtual-transition remains below 3\% over the [150--350]~meV window [Fig.~\ref{Fig.1}(d)  and SM Fig. S4].}

\addtext{Because the approximation holds over such a broad window ($\sim$200~meV), strained TBG is well suited for testing the IC--SC connection experimentally. First, the CIC and LSC conductivities are measured at the same photon energy. Second, from these two measured conductivities alone, our relation yields an inferred relaxation time based on Eq.~(8):}
\begin{equation}
\tau_{\text{inf}}  \equiv \frac{\sigma^{cba}_{\text{CIC}}+\sigma^{bca}_{\text{CIC}}}{2\omega\,\sigma^{abc}_{\text{LSC}}}.
\end{equation}
\addtext{Third, $\tau_{\text{inf}} $ is compared with the relaxation time $\tau$ of the optically injected current obtained by an independent method under matched conditions.}

To test the connection between LIC and CSC, ferromagnetic, antiferromagnetic, and altermagnetic materials that host approximately linear Dirac/Weyl dispersions are suitable candidates. As a representative example, we consider MnGeO$_3$ (SM Sec.  VIII), which has been predicted to be an antiferromagnetic Dirac semimetal that preserves $\mathcal{PT}$ symmetry \cite{MNG1,MNG2}   [Fig. \ref{Fig.1}(e)]. In such $\mathcal{PT}$-symmetric systems,  $\sigma_{\mathrm{CIC}}$ and $\sigma_{\mathrm{LSC}}$ are symmetry-forbidden,  which makes MnGeO$_3$ an ideal platform for verifying the connection between $\sigma_{\mathrm{LIC}}$ and $\sigma_{\mathrm{CSC}}$, as illustrated in Fig. \ref{Fig.1}(f).

Previous studies have connected IC to asymmetries of quantum geometry in momentum space \cite{Nagaosa17,NH-08,Ma2021Topology,Bao2022,NH-09}. Here we go a step further and show that injection and shift currents are directly linked through an interband quantum-geometric dipole. Specifically, the CIC and the LSC are related by the interband Berry-curvature dipole $D^{a;bc}$, whereas the LIC and the CSC are linked through the interband quantum-metric dipole $Q^{a;bc}$ (SM Sec.  V):
	\begin{equation}
		\begin{split}
			D^{a;bc}(\omega) \equiv \int_{\mathbf{k}} \sum_{n,m}f_{nm}\,\partial_{k_a}\Omega^{bc}_{nm}\,\Theta(\omega-\omega_{mn}),\\
			Q^{a;bc}(\omega) \equiv \int_{\mathbf{k}} \sum_{n,m}f_{nm}\,\partial_{k_a} g^{bc}_{nm}\,
			\Theta(\omega-\omega_{mn}),
		\end{split}\label{eq:0002}
	\end{equation}
where
$\Omega^{bc}_{nm}$ and $g^{bc}_{nm}$ denote the interband Berry curvature and the interband quantum metric, respectively, and $\Theta$ is the Heaviside step function (see SM, Sec. V).	Collecting these results, the nonlinear optical conductivities can be written in a unified interband quantum-geometric form:
	\begin{equation}
		\begin{split}
			\sigma_{\mathrm{CIC}}^{abc}
			&=
			-\tau\frac{\pi e^{3}}{\hbar^{2}}\,D^{a;bc}, 
			  \;\;\;\;\;\;
			\sigma_{\mathrm{LSC}}^{abc}
			\approx
			-\frac{\pi e^{3}}{2\hbar^{2}\omega}
			\left(D^{c;ba}+D^{b;ca}\right),\\
			\sigma_{\mathrm{LIC}}^{abc}
			&=
			+2\tau\frac{\pi e^{3}}{\hbar^{2}}\,Q^{a;bc}, \;\;\;\;
			\sigma_{\mathrm{CSC}}^{abc}
			\approx
			-\frac{\pi e^{3}}{\hbar^{2}\omega}
			\left(Q^{c;ba}-Q^{b;ca}\right),
		\end{split}
		\label{eq:12}
	\end{equation}
	Equation \eqref{eq:12} shows that IC and SC are effectively equivalent within the linearized band approximation in the low-frequency limit, differing only in how the quantum-geometric dipole is projected. For the IC, the dipole is aligned with the current direction  [Fig. \ref{Fig.1}(g)-Left], while for the SC, the dipole moments are closely related to the polarization direction of the laser. For instance, $\sigma^{zxy}_{\text{LSC}}$ measures the Berry-curvature dipoles along the $x$ and $y$ directions  [Fig. \ref{Fig.1}(g)-Right]. A complete correspondence between the SC conductivity tensor and the quantum-geometric dipoles is provided in SM Sec.  V.

\begin{figure} 
	\includegraphics[width=\linewidth]{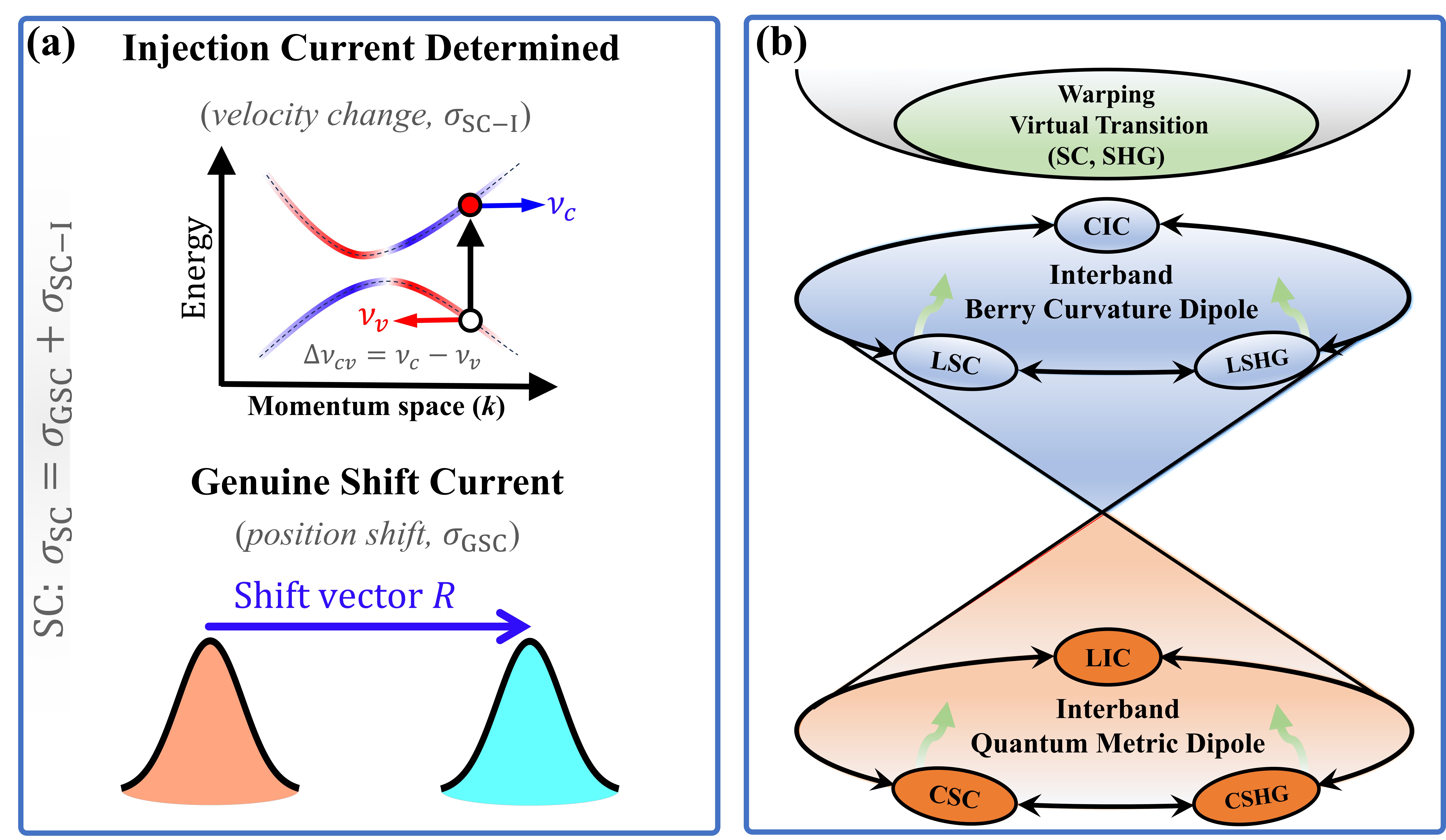}
	\caption{	\addtext{ {\bf Genuine SC and the extension to SHG}   (a)  Real-space decomposition of the SC into its IC-determined and genuine sectors, $\sigma_{\mathrm{SC}} = \sigma_{\mathrm{GSC}} + \sigma_{\text{SC-I}}$ [Eq.~(\ref{eq:GSC})]. Top: the IC-determined sector $\sigma_{\text{SC-}I}$ arises from the group-velocity mismatch $\Delta v_{cv}=v_c-v_v$ between the valence and conduction states connected by the photon-induced transition $\hbar\omega$ --- the same momentum-space asymmetry that generates the IC. Bottom: the genuine SC $\sigma_{\mathrm{GSC}}$ is purely positional, arising instead from a real-space displacement of the electronic wave packet by the shift vector $R$, with no accompanying velocity asymmetry.  (b) Within a linearized band approximation in the low-frequency limit, the upper (blue) sector shows Berry-curvature-dipole driven responses linking the CIC with the LSC and LSHG, while the lower (orange) sector shows quantum-metric-dipole driven responses connecting the LIC with the CSC and CSHG. In the general case, warping and virtual-transition contributions must be included, giving rise to extra contributions to the SC and SHG.}}
	\label{Fig.2}
\end{figure} 

\addtext{\textbf{Beyond the linear limit: the genuine shift current.}} \addtext{We now consider the full identities, where the second-derivative and virtual-transition sectors can be significant. The conventional distinction between IC and SC concerns velocity asymmetry and wave-packet displacement, respectively ~\cite{NH-09}. An optical transition contributing to SC can also involve a change in group velocity between the resonant bands.}

\addtext{Our decomposition makes the group-velocity contribution to the SC explicit. The terms containing the group-velocity difference of the resonant band pair are collected in $\sigma_{\text{LSC-I}}^{abc}=\frac{1}{2\tau\omega}\left(\sigma^{cba}_{\text{CIC}}+\sigma^{bca}_{\text{CIC}}\right)$ and $\sigma_{\text{CSC-I}}^{abc}=-\frac{1}{2\tau\omega}\left(\sigma^{cba}_{\text{LIC}}-\sigma^{bca}_{\text{LIC}}\right)$ for the LSC and CSC, respectively. Removing this IC-determined sector leaves  (SM Sec.~VI.A)}
\begin{equation}
\sigma_{\text{GSC}}^{abc}\equiv\sigma_{\text{SC}}^{abc}-\sigma_{\text{SC-I}}^{abc}=\sigma_{\text{SC}-\mathcal{W}}^{abc}+\sigma_{\text{SC}-\mathcal{V}}^{abc}.
\label{eq:GSC}
\end{equation}

\addtext{We refer to this remainder as the genuine SC. Here ``genuine'' specifies the residual sector of this decomposition: it contains no explicit group-velocity-difference terms of the resonant band pair, and it contains only the wave-packet-displacement response that remains once the injection-related contribution has been removed  [Fig. \ref{Fig.2}(a)].}
 
\addtext{Whenever the IC-determined projection vanishes, the measured SC is entirely the genuine SC. For the LSC in nonmagnetic materials, the condition is $\sigma^{cba}_{\text{CIC}}+\sigma^{bca}_{\text{CIC}}=0$. The diagonal components $xxx$, $yyy$, and $zzz$, when symmetry-allowed, satisfy this condition because $\sigma^{aaa}_{\text{CIC}}=0$ by antisymmetry in the optical indices; the component along the polar axis of a polar crystal is one example. Crystal symmetry provides further opportunities. In the point groups $C_{3h}$, $D_{3h}$, $T$, and $T_{d}$, the IC projection vanishes for every index combination, so every symmetry-allowed LSC component is a genuine SC. Other groups permit selected components for which the relevant CIC elements vanish separately or cancel, whereas in groups $D_{2}$, $S_{4}$, $D_{4}$, $D_{2d}$, and $D_{6}$ all allowed components contain an injection-related contribution and the genuine SC must be obtained by subtraction; the LSC vanishes identically in group $O$. SM Table S3 lists, for the 21 noncentrosymmetric point groups of nonmagnetic crystals, the symmetry-allowed LSC components that are free of the injection-related sector, providing a practical guide to selecting current and polarization configurations for a direct measurement of the genuine SC.}

\addtext{The two genuine SC components have distinct microscopic origins~\cite{Sipe2000, Nagaosa2020Low}. The $\sigma_{\text{SC}-\mathcal{W}}$ contains the second-derivative vertex $\partial_a\partial_b H$, whereas the $\sigma_{\text{SC}-\mathcal{V}}$ arises from virtual transitions through additional bands. Despite their importance in real materials~\cite{SC2,Nagaosa2020Low}, the $\sigma_{\text{SC}-\mathcal{W}}$ and $\sigma_{\text{SC}-\mathcal{V}}$ sectors have generally not been separable in a measurement of the SC response. Our identities provide a route to this separation under controlled conditions (SM Sec.~VI.B).}

\addtext{We also note that, within the clean limit, assuming an independent-particle interband framework, the quantum-geometric-dipole contribution to the SC contains an explicit $1/\omega$ prefactor. Related low-frequency divergences were identified in topological semimetals \cite{Nagaosa2020Low}. Our formulation suggests that such enhancement may also occur in other metals, provided  the relevant interband quantum-geometric dipole projection approaches a finite nonzero value. In real materials, however, finite scattering is expected to cut off this clean limit enhancement at sufficiently low frequencies.}

\addtext{\textbf{Extension to second-harmonic generation (SHG).}} \addtext{We next apply our approach to consider SHG to obtain relations between the SHG and SC conductivities, with explicit second-derivative and virtual-transition remainders  (SM Sec.~VII),}
\begin{equation}
	\begin{aligned}
		\sigma^{abc}_{\mathrm{LSHG}}(\omega,2\omega)
		=&
		\frac{1}{4\tau\omega} \left[\sigma^{cba}_{\text{CIC}}(\omega) + \sigma^{b ca}_{\text{CIC}}(\omega) \right]\\& -\frac{1}{\tau\omega} \left[\sigma^{cba}_{\text{CIC}}(2\omega) + \sigma^{b ca}_{\text{CIC}}(2\omega) \right]\\
		&+\sigma^{abc}_{\mathrm{LSHG\text{-}\mathcal W}}(\omega,2\omega)+\sigma^{abc}_{\mathrm{LSHG\text{-}\mathcal V}}(\omega,2\omega),
	\end{aligned}
	\label{eq:SHG_full_L_gen}
\end{equation}
\begin{equation}
	\begin{aligned}
		\sigma^{abc}_{\mathrm{CSHG}}(\omega,2\omega)
		=&
		\frac{-1}{4\tau\omega} \left[\sigma^{cba}_{\text{LIC}}(\omega)- \sigma^{b ca}_{\text{LIC}}(\omega) \right] \\& +\frac{1}{\tau\omega} \left[\sigma^{cba}_{\text{LIC}}(2\omega) - \sigma^{b ca}_{\text{LIC}}(2\omega) \right]\\
		&+\sigma^{abc}_{\mathrm{CSHG\text{-}\mathcal W}}(\omega,2\omega)+\sigma^{abc}_{\mathrm{CSHG\text{-}\mathcal V}}(\omega,2\omega),
	\end{aligned}
	\label{eq:SHG_full_C_gen}
\end{equation}
\addtext{Quantum-geometric decompositions of resonant SHG have been discussed in the literature~\cite{Morimoto16,Ahn2021,Bhalla2022}. The key new insight here is a response-level identity that ties the SHG conductivity to independently measurable photocurrents [Fig. \ref{Fig.2}(b)]. Our IC--SC connection is thus not an isolated algebraic coincidence but one member of a broader class of cross-response relations generated by the same underlying sum-rule structure, which suggests a systematic route to hidden quantitative relations among the various nonlinear optical responses.}

\addtext{In summary, we have uncovered a hidden connection between IC and SC, demonstrating that distinct microscopic origins and geometric descriptions need not preclude quantitative connections between nonlinear optical responses. Such connections complement the geometric characterization of individual responses by revealing how they constrain one another and how their contributions can be disentangled experimentally. The extension from photocurrents to second-harmonic generation suggests that the connection established here is part of a broader network of relations among various nonlinear optical phenomena. Identifying the scope and structure of this network offers a new pathway for a deeper understanding of nonlinear optical phenomena in materials.}


\end{document}